\documentclass[journal]{IEEEtran}
\usepackage[cmex10]{amsmath}

\ifCLASSINFOpdf

\else

  \usepackage[dvips]{graphicx}

\fi

\begin{document}

\title{Cascade-gain-switching for generating 3.5-$\mu$m nanosecond pulses from low-cost monolithic fiber lasers}

\author{Jianlong Yang, Haizhe Zhong, Yulong Tang, Shuaiyi Zhang, and Dianyuan Fan
% <-this % stops a space
\thanks{This work was supported in part by the Natural Science Foundation of China (NSFC) (No. 61505113 and No. 61405101); Science and Technology Project of Shenzhen (No. JCYJ20160308091733202 and No. ZDSYS201707271014468); Science and Technology Planning Project of Guangdong Province (No. 2016B050501005); Educational Commission of Guangdong Province (No. 2016KCXTD006). Corresponding author: Haizhe Zhong (e-mail: haizhe.zhong@szu.edu.cn)}

\thanks{J. Yang, H. Zhong, and D. Fan are with Engineering Technology Research Center for 2D Material Information Function Devices and Systems of Guangdong Province, Key Laboratory of Optoelectronic Devices and Systems of Ministry of Education and Guangdong Province, College of Optoelectronic Engineering, Shenzhen University, Shenzhen 518060, China.}
\thanks{Y. Tang is with Key Laboratory for Laser Plasma (Ministry of Education), School of Physics and Astronomy, Collaborative Innovation Center of IFSA (CICIFSA), Shanghai Jiao Tong University, Shanghai 200240, China.} 
\thanks{S. Zhang is with School of Mathematics and Physics, Qingdao University of Science \& Technology, Qingdao 266061, China.}% <-this % stops a space
%\thanks{Manuscript received April 19, 2005; revised December 27, 2012.}}
}

\maketitle
\begin{abstract}
We propose a novel laser configuration that can output 3.5-$\mu$m nanosecond laser pulses based on a simple and monolithic fiber structure. Cascade-gain-switching, which converts the wavelength of nanosecond pulses from 1.55 $\mu$m to 3.5 $\mu$m by two successive gain-switching processes. Instead of using expensive pump sources at special wavelengths and bulky active or passive modulation elements for Q-switching or mode-locking, the cascade gain-switching only requires the pumping of an electric-modulated 1.55-$\mu$m pulsed laser and two continuous-wave (CW) 975-nm laser diodes. They are all standard products for fiber optic communication applications, which can greatly lower the cost of mid-infrared laser pulse generation. To investigate the feasibility of this configuration, we numerically simulated the cascade-gain-switching processes by comprehensive rate-equation models. In single-shot regime, for stable 3.5-$\mu$m pulsed lasing, the CW 975-nm pump should be turned on at least $\sim500$ $\mu$s ahead of the 1.55-$\mu$m pulsed pump. It shows that the pulse width of the 1.55-$\mu$m pump has major impact on the temporal shape of the intermediate 1.97-$\mu$m pulse while has neglected influence on the generated 3.5-$\mu$m pulse. On the other hand, increasing the CW pump power can significantly improve the output peak power and shorten the pulse when the pump power is less than $\sim4$ W. In the repetitive-pulse regime, we found the 3.5-$\mu$m pulse train can be stably outputted when the repetition rate is $<=100$ kHz. As the repetition rate increases, the duration of reaching the stable operation increases. When the repetition rate is large than 100 kHz, the stable operation cannot be established because the rate of consuming the population on the $^4I_{11/2}$ level of  $Er^{3+}$ ions is faster than the rate of building the population. 
\end{abstract}
\IEEEpeerreviewmaketitle
\section{Introduction}
Nanosecond mid-infrared lasers are of significant interests in the applications such as non-metallic material marking, environmental sensing, and biomolecular detection \cite{rudy14}. In recent years, as the development of laser materials, pump techniques, and optical components, the reported performances of mid-infrared nanosecon lasers have been progressing rapidly (see \cite{np12} and reference therein).\\
\indent Many approaches can be employed to generate mid-infrared nanosecond lasers. It usually includes rare-earth-doped solid-state and fiber lasers, quantum cascade lasers, nonlinear wavelength conversion and so on \cite{rudy14}. Among them, fiber lasers have the advantages of structure simplicity, compactness, near-diffraction-limited beam quality, and high efficiency \cite{zhu17}. These attractive characteristics has greatly promoted the development of mid-infrared nanosecond fiber lasers. In 2011, Tokita \textit{et al.} realized an active Q-switched Er:ZBLAN fiber laser outputting 2.8-$\mu$m pulses with the average power up to 12 W and pulse duration down to 90 ns at a repetition rate of 120 kHz \cite{tokita11}. Gorjan \textit{et al.} built a gain-switched Er:ZBLAN fiber laser pumped by a pulsed diode system and achieved a pulse width of $\sim300$ ns and the peak power up to 68 W at 2.8 $\mu$m \cite{gorjan11}. In 2012, Li \textit{et al.} reported a Q-switched Ho-doped fluoride fiber laser that could deliver a pulse energy of 29 $\mu$J and a pulse width of 380 ns at 3 $\mu$m \cite{li12}. Hu \textit{et al.} developed a Q-switched Ho/Pr-codoped fiber laser and acquired a pulse width of 78 ns and a peak power of 77 W at 2.9 $\mu$m \cite{hu12}. It should be noted that the laser wavelengths of these works are limited at around 3 $\mu$m. The development of nanosecond lasers with even longer wavelengths will have both research and application interests.\\
\indent In 2014, Henderson-Sapir \textit{et al.} developed a dual-wavelength pumping (DWP) strategy for Er:ZBLAN that could improve the 3.5-$\mu$m laser generation efficiency significantly \cite{hs14}. Instead of using the pump at $\sim650$ nm, this method proposed to use two pump sources at 975 nm and 1.97 $\mu$m, respectively.  Very recently, Maes \textit{et al.} implemented this method in a monolithic fiber structure and achieved 5.6-W continuous-wave (CW) output at 3.55 $\mu$m \cite{maes17}. This remarkable result further verifies the effectiveness of this DWP approach. In \cite{malouf16}, numerical model of the DWP Er:ZBLAN fiber lasers was established. Good accordance between calculation and previous experimental results was achieved. They further numerically extended the application of the DWP to Quasi-CW (QCW) regime, which could generate QCW pulses (several-hundred-microsecond with leading-edge spikes) at 1-kHz. However, to the best of our knowledge, this approach has not been used in pulsed regime to generate 3.5-$\mu$m pulses shorter than microsecond.\\
\indent In this paper, we propose to use cascade-gain-switching technique to achieve 3.5-$\mu$m nanosecond pulses. The CW pump at $\sim975$ nm could store pump energy in $^4I_{11/2}$ level thus decrease the pulse lasing threshold. The 1.97-$\mu$m nanosecond pump, on the other hand,  can be used to realize quasi-in-band gain-switching between $^4I_{11/2}$ and $^4F_{9/2}$ levels. Because the in-band-pumped gain-switched Tm-doped fiber lasers are one of the most robust and simple approaches to generate 2-$\mu$m nanosecond pulses \cite{yang13}, if we use them as the pulsed pump, it can form a cascade gain-switching configuration: switches the wavelength of nanosecond pulses from 1.55 $\mu$m to 1.97 $\mu$m to 3.5 $\mu$m.
\section{Methods}
\subsection{Cascade-gain-switching scheme}
\begin{figure}[hhh]
\centering
\includegraphics[width=8cm]{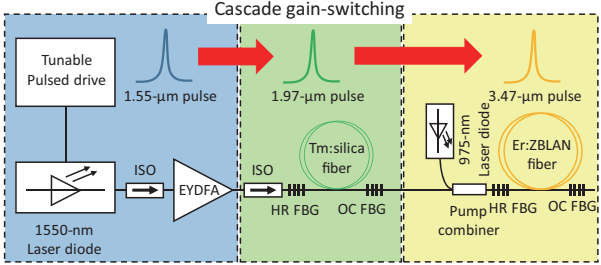}
\caption{Illustration of the configuration of cascade-gain-switching fiber lasers. EYDFA: Er/Yb-codoped fiber amplifier; ISO: fiber isolator; FBG: fiber Bragg grating; HR: Highly-reflective mirror; OC: output coupler.}
\end{figure}
We illustrated the configuration of the cascade-gain-switching fiber lasers in Fig.~1. The generation of the 1.55-$\mu$m pulses can take advantage of the highly-developed telecommunication fiber sources and components. A 1550-nm laser diode is driven by a tunable pulse generation circuit. The output pulse width and repetition rate can be adjusted directly. The optical power output from a laser diode is usually limited, so a Er/Yb-codoped fiber amplifier (EYDFA) is included to boost the peak power of the 1.55-$\mu$m pulses. Benefited from the high optical gain ($>$ 20 dB) provided by this amplifier, further amplification is usually unnecessary for the gain-switching from 1.55 $\mu$m to 1.97 $\mu$m. To prevent parasitic lasing, two fiber isolators are inserted before and after the EYDFA. Then the 1.55-$\mu$m pulses are injected into a Tm:silica fiber oscillator constituting of a section of Tm-doped fiber and two fiber Bragg gratings (FBGs) as highly-reflective mirror and output coupler, respectively. The generated 1.97-$\mu$m pulses combined with the 975-nm CW light from a high-power laser diode are injected into a Er:ZBLAN fiber oscillator through a (N+1)$\times$1 pump combiner. The 975-nm pump light is multi-mode thus going through one pump arm of the pump combiner. The 1.97-$\mu$m pulses are close to single-mode so going through the signal arm of the pump combiner. Because the attenuation of silica material will dramatically increase when the wavelength longer than 2.3 $\mu$m, so no isolator is required between these two oscillators. The Er:ZBLAN fiber oscillator and two FBGs centerd at 3.47 $\mu$m will form the gain-switched 3.47 $\mu$m pulses.\\
\subsection{Model of gain-switched in-band-pumped thulium-doped fiber lasers}
The right side of Fig. 2 shows the simplified energy diagram of Tm:silica fiber for the in-band-pumped gain-switching operation, which only involves two energy levels $^3H_{6}$ and $^3F_{4}$. Compared with the lasing process having the relaxation from pump absorption to laser radiation, the in-band pumping has instinct advantages on pulse regulation \cite{yang13}. This technique has been established for 10 years \cite{jiang07} and used in many applications such as mid-infrared optical parametric oscillator \cite{creeden08}, mid-infrared supercontinuum generation \cite{swiderski13}, and high-power fiber amplification \cite{yang15,yang132}. In our previous publication \cite{yang14}, the theoretical model of this kind of lasers has been described in very detail, so we will not repeat it here. Besides, a section of passive fiber is also included in this model to temporally stretch the pulse to avoid potential nonlinear effects during the laser oscillation and pulse amplification. The equation for the passive propagation is:
\begin{equation}
\pm\frac{\partial P^{\pm}(z,t)}{\partial z}+\frac{\partial P^{\pm}(z,t)}{v_p\partial t}=-\alpha P^{\pm}(z,t)
\end{equation}
where $P$ is the optical power of the pulse. $z$ is the axial position along the gain fiber from 0 to $l$, which is the length of the fiber. $+$ is the direction from 0 to $l$ and $-$ is the opposite. $t$ is time. $\alpha$ is the fiber attenuation.
\subsection{Model of hybrid-pumped gain-switched Er:ZBLAN fiber lasers}
\begin{figure*}[htb]
\centering
\includegraphics[width=16cm]{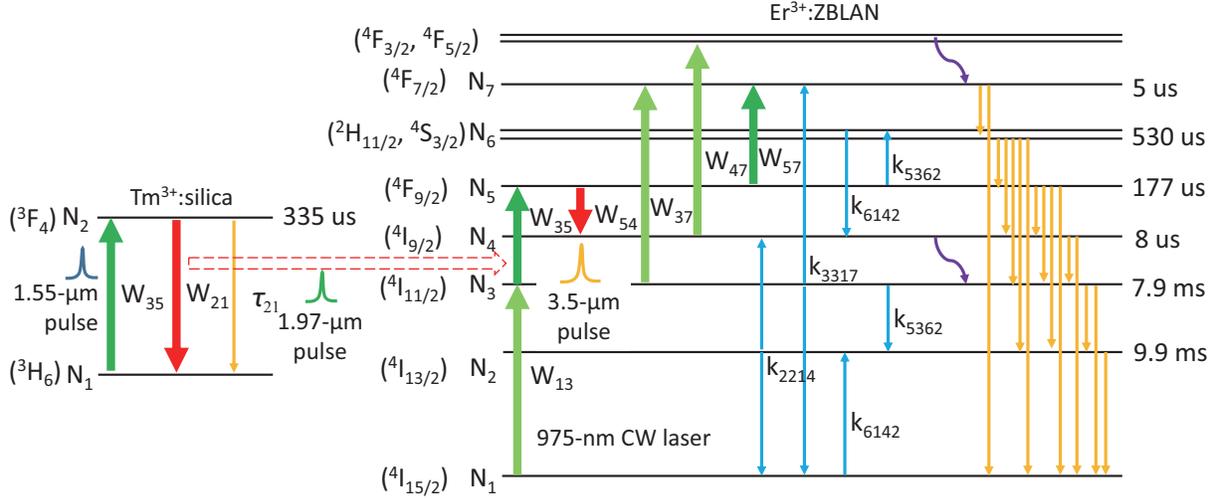}
\caption{Simplified energy diagrams of Tm:silica fiber and Er:ZBLAN fiber and the involved transitions among energy levels using the cascade-gain-switching scheme.}
\end{figure*}
The left side of Fig. 2 shows the energy diagram of Er:ZBLAN fiber and the major involved transitions among energy levels using the hybrid-pumping scheme. For clarity, we did not plot the corresponding stimulated emission or absorption processes related to the pump and laser transitions and excited state absorptions (ESAs). But they all have been included in the model. We also included the newly-identified upconversion process reported in \cite{hs16} and all the other energy transition processes in the previous theoretical works about the DWP \cite{malouf16,maes172}. In this figure, $N_1$ to $N_7$ are the populations of the energy level $^4I_{15/2}$ to $^4F_{7/2}$, respectively. $W_{13}$ is stimulated absorption induced by the 975-nm CW pump. $W_{35}$ is stimulated absorption induced by the 1.97-$\mu$m pulse pump. $W_{35}$ is the 3.5-$\mu$m laser emission.  $W_{37}$, $W_{47}$, and $W_{57}$ are ESAs. $k_{6142}$ is a cross relaxation process. $k_{2214}$, $k_{3317}$, and $k_{5362}$ are energy upconversion processes. Two multi-photon decay processes are indicated by purple-color arrows. The branch ratios of single-photon decays are labeled by $\tau_{ij}$, where $i$ and $j$ represent the corresponding energy levels. The life times of the excited states from \cite{malouf16} are listed in the left side of the diagram. As shown in this figure, the hybrid pumping can be illustrated as: the CW pump at $\sim975$ nm will store pump energy in $^4I_{11/2}$ level thus decrease the pulse lasing threshold. The 1.97-$\mu$m nanosecond pump will induce a burst of energy transition from $^4I_{11/2}$ to $^4F_{9/2}$ thus lead to the lasing from $^4F_{9/2}$ to $^4I_{9/2}$. Because the life time of $^4I_{9/2}$ is much shorter than $^4I_{11/2}$, it can be regarded as a quasi-in-band gain-switching process.\\
\indent Limited by the length of this paper and also because they are quite similar to those in the previous publications \cite{maes172,malouf16}, we will not list the rate equations of all the energy levels here. On the other hand, we demonstrate the optical power propagation equations below because we are using a time-dependent model instead of the steady-state models in the previous papers.
\begin{equation}
\begin{split}
\pm \frac{\partial P^{\pm}_{975}(z,t)}{\partial z}&+\frac{\partial P^{\pm}_{975}(z,t)}{v_g\partial t}=\Gamma_p^{clad}(-\sigma_{13}N_1+\sigma_{31}N_3\\&-\sigma_{37}N_3+\sigma_{73}N_7-\sigma_{47}N_4+\\&\sigma_{74}N_7)P^{\pm}_{975}(z,t)-\alpha P^{\pm}_{975}(z,t),
\end{split}
\end{equation}
\begin{equation}
\begin{split}
\pm \frac{\partial P^{\pm}_{1970}(z,t)}{\partial z}&+\frac{\partial P^{\pm}_{1970}(z,t)}{v_g\partial t}=\Gamma_p^{core}(-\sigma_{35}N_3+\\&\sigma_{53}N_5-\sigma_{57}N_5)P^{\pm}_{1970}(z,t)-\alpha P^{\pm}_{1970}(z,t),
\end{split}
\end{equation}
\begin{equation}
\begin{split}
\pm \frac{\partial S^{\pm}_{3470}(z,t)}{\partial z}&+\frac{\partial S^{\pm}_{3470}(z,t)}{v_g\partial t}=\Gamma_s^{core}(-\sigma_{45}N_4+\\&\sigma_{54}N_5)S^{\pm}_{3470}(z,t)-\alpha S^{\pm}_{3470}(z,t)+\\&\Gamma_s^{core}\frac{2hc^2}{\lambda_s^3}\sigma_{54}N_5\Delta\lambda_{ASE}.
\end{split}
\end{equation}
Equation (2), (3), and (4) are the optical power propagation equations of the 975-nm pump $P_{975}$, 1970-nm pump $P_{1970}$ and 3470-nm laser $S_{3470}$, respectively. $\sigma_{ij}$ are the absorption and emission cross sections. $\Gamma_p^{clad}$, $\Gamma_p^{core}$, and $\Gamma_s^{core}$ are their corresponding power filling as defined in \cite{yang14}. The last term in Equation (4) is the spontaneous emission term as the seed of the laser oscillator. $h$ is Planck constant. $c$ is the speed of light in vacuum. $\Delta \lambda_{ASE}$ is the bandwidth of the amplified spontaneous emission (ASE).\\
\indent The boundary conditions to solve these equations are:
\begin{eqnarray}
P_{975}(z=0,t)=P_{LD},\\
P_{1970}(z=0,t)=P_{p0}(t),\\
S_{3470}^+(z=0,t)=R_{HR}S_{3470}^-(z=0,t),\\
S_{3470}^-(z=l,t)=R_{OC}S_{3470}^+(z=l,t),
\end{eqnarray}
where $P_{LD}$ is the CW power outputted from the 975-nm laser diode. $P_{p0}(t)$ is the injected 1970-nm gain-switched laser pulses. $R_{HR}$ and $R_{OC}$ are the reflectivity of the HR FBG and OC FBG, respectively.
\subsection{Physical parameters and numerical method}
\begin{table}
\caption{Physical parameters employed in the numerical simulation}
\begin{center}
\begin{tabular}{p{2.5cm}p{1.5cm}p{2.5cm}p{1cm}}
\hline
Parameter & Unit & Value & Ref.\\
\hline
Er$^{3+}$ concentration & m$^{-3}$ & $1.6\times10^{26}$ & \cite{maes17}\\
Core diameter & $\mu$m & $16.5$ & \cite{maes17} \\
Cladding diameter & $\mu$m & $170$ & \cite{maes17} \\
$l$ & m & $3.4$ & \cite{maes17}\\
$R_{HR}$ & \% & 99 & \cite{malouf16}\\
$R_{OC}$ & \% & 55 & \cite{maes17}\\ 
$\alpha$ & m$^{-1}$ & 0.035& \cite{maes17,malouf16}\\
$\sigma_{13}$,$\sigma_{31}$,$\sigma_{35}$,$\sigma_{53}$ & $10^{-26}$ m$^2$ & 3.77,1.64,30,37.5& \cite{maes17}\\
$\sigma_{45}$,$\sigma_{54}$ & $10^{-26}$ m$^2$ & 1.5,1.08 & \cite{maes17}\\
$\sigma_{37}$,$\sigma_{73}$,$\sigma_{47}$,$\sigma_{74}$,$\sigma_{57}$ & $10^{-26}$ m$^2$ & 26.4,31.9,13.5,17.4,0.7& \cite{maes17,malouf16}\\
$k_{2214}$,$k_{3317}$,$k_{6142}$,$_{5362}$ & $10^{-23}$ m$^3$s$^{-1}$ & 1.3,0.16,0.48,2.5& \cite{maes17}\\
$\tau_{76}$,$\tau_{71}$& - & 0.99,0.01 & \cite{malouf16}\\
$\tau_{65}$,$\tau_{64}$,$\tau_{63}$,$\tau_{62}$,$\tau_{61}$& - & 0.285, 0.029, 0.014, 0.193, 0.479 & \cite{malouf16}\\
$\tau_{54}$,$\tau_{53}$,$\tau_{52}$,$\tau_{51}$& - & 0.808, 0.008, 0.009, 0.175 & \cite{malouf16}\\
$\tau_{43}$,$\tau_{41}$& - & 0.999, 0.001 & \cite{malouf16}\\
$\tau_{32}$,$\tau_{31}$& - & 0.182, 0.818 & \cite{malouf16}\\
$\tau_{21}$& - & 1 & \cite{malouf16}\\
$\Delta\lambda_{ASE}$ & nm & 15 & -\\
\hline
\end{tabular}
\end{center}
\end{table}
Realistic fiber and optical parameters were employed to solve the theoretical model above as listed in Table 1. The ASE bandwidth $\Delta\lambda_{ASE}$ is simply for the initialization of the laser oscillation. We have verified changing this value would not influence the results. The numerical method we used is finite difference in time-domain (FDTD) algorithm. The optical power at a specific moment and position can be determined by the populations of the energy levels at this moment and position, the optical power at this moment but in the previous position, and the optical power at this position but in the previous moment. Such calculations proceed the axial directions first from 0 to $l$ then $l$ to 0, iteratively. the convergence is achieved when the power difference of two successive iterations is less than 0.1\%. This method has been successfully applied in the simulations of different types of fiber lasers and amplifiers \cite{yang12,yang152,yang16} and achieved excellent accordance with experiments. 
\begin{figure*}[!!!htb]
\centering
\includegraphics[width=14cm]{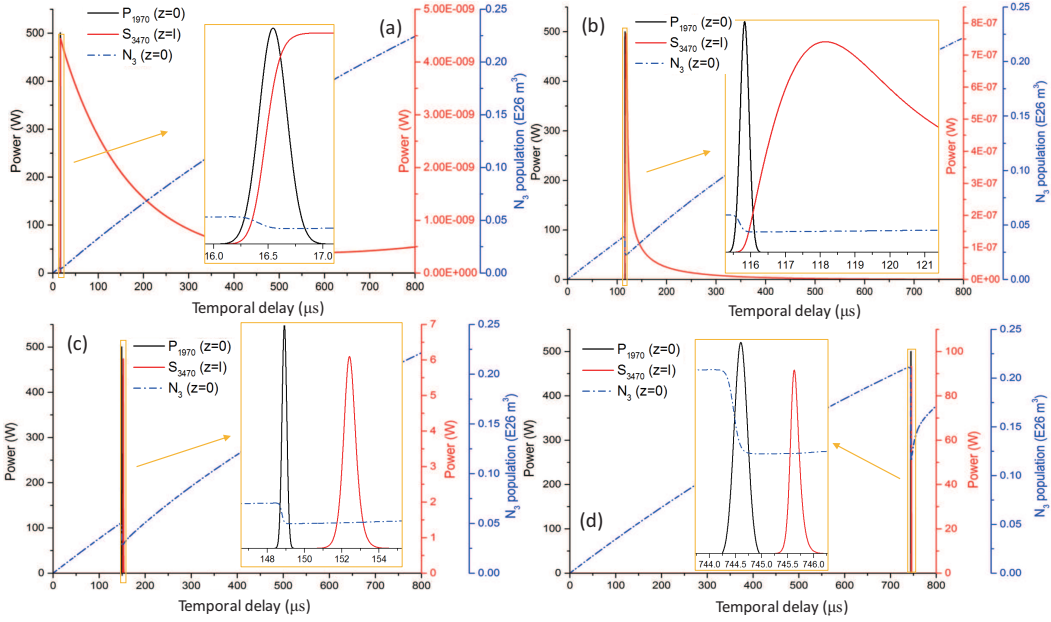}
\caption{Temporal evolution of the 3.5-$\mu$m output $S_{3470}(z=l)$ and the CW pump induced population accumulation $N_3(z=0)$ at different pulse delays: (a) 17 $\mu$s; (b)115 $\mu$s; (c) 149 $\mu$s; (d) 745 $\mu$s. Pump pulse $P_{1970}(z=0)$ is also plotted as a reference. The window inside is a zoomed view.}
\end{figure*}
\section{Results}
\subsection{Lasing dynamics of gain-switched 3.5-$\mu$m pulses}
In \cite{malouf16}, 1-kHz pulse train with a pulse width of 300 $\mu$s at 1.97 $\mu$m was employed to generate QCW pulses at 2.8 $\mu$m and 3.5 $\mu$m. The pulse widths of the generated 2.8-$\mu$m and 3.5-$\mu$m lasers are $\sim$600 $\mu$s and 150 $\mu$s, respectively. These QCW pulses have strong relaxation oscillation on the temporal envelope and the lasing dynamics is very similar with the situation in CW regime. Gain-switching, on the other hand, operates at a much shorter temporal scale. The pump timing and power should be in well control to lase only the first spike of the relaxation oscillation. Here we investigate the lasing dynamics of the 3.5-$\mu$m gain-switched pulses.  We employed single 1970-nm pump pulse to isolate the influence of repetition rate. Based on the previous works on the gain-switched Tm-doped fiber lasers \cite{jiang07,yang13,yang14}, it is reasonable to assume the 1970-nm pump pulse has a Gaussian profile. We set the pump pulse width to 300 ns  because it can be easily acquired in the experiments without the optimization of the cavity length \cite{yang13}. The CW pump peak power was set to 5 W unless state otherwise.\\
\indent Figure 3 shows the temporal evolution of the 3.5-$\mu$m output $S_{3470}(z=l)$ and the CW pump induced population accumulation $N_3(z=0)$ at different pulse delays. We set a peak power of 0.5 kW for the pump pulse at 1970 nm and a CW power of 4.5 W at 975 nm. The pump pulse was also plotted as a reference. 
\indent In Fig.~3(a), when the delay is 17 $\mu$s, the accumulated population is at a very low level, only nW-level ASE is emitted at 3470 nm. The pump pulse consumes a small amount of accumulated energy. When the delay increases to 115 $\mu$s as shown in Fig.~3(b), the laser still not forms but the ASE power has increased by two orders of magnitude. When the delay further increases to 149 $\mu$s [Fig.~3(c)], laser pulse forms. However, the 3470-nm pulse width is much wider than the pump and the peak power is very low. In Fig.~3(d), at a temporal delay of 745 $\mu$s, the accumulated population has been more than 20\% of the concentration of the Er$^{3+}$. The 3470-nm laser pulse becomes narrower than the pump pulse and the peak power is close to 100 W. At this time, the population consumption becomes significant, which takes around 40\% of the total accumulated population.\\
\begin{figure}[htb]
\centering
\includegraphics[width=8cm]{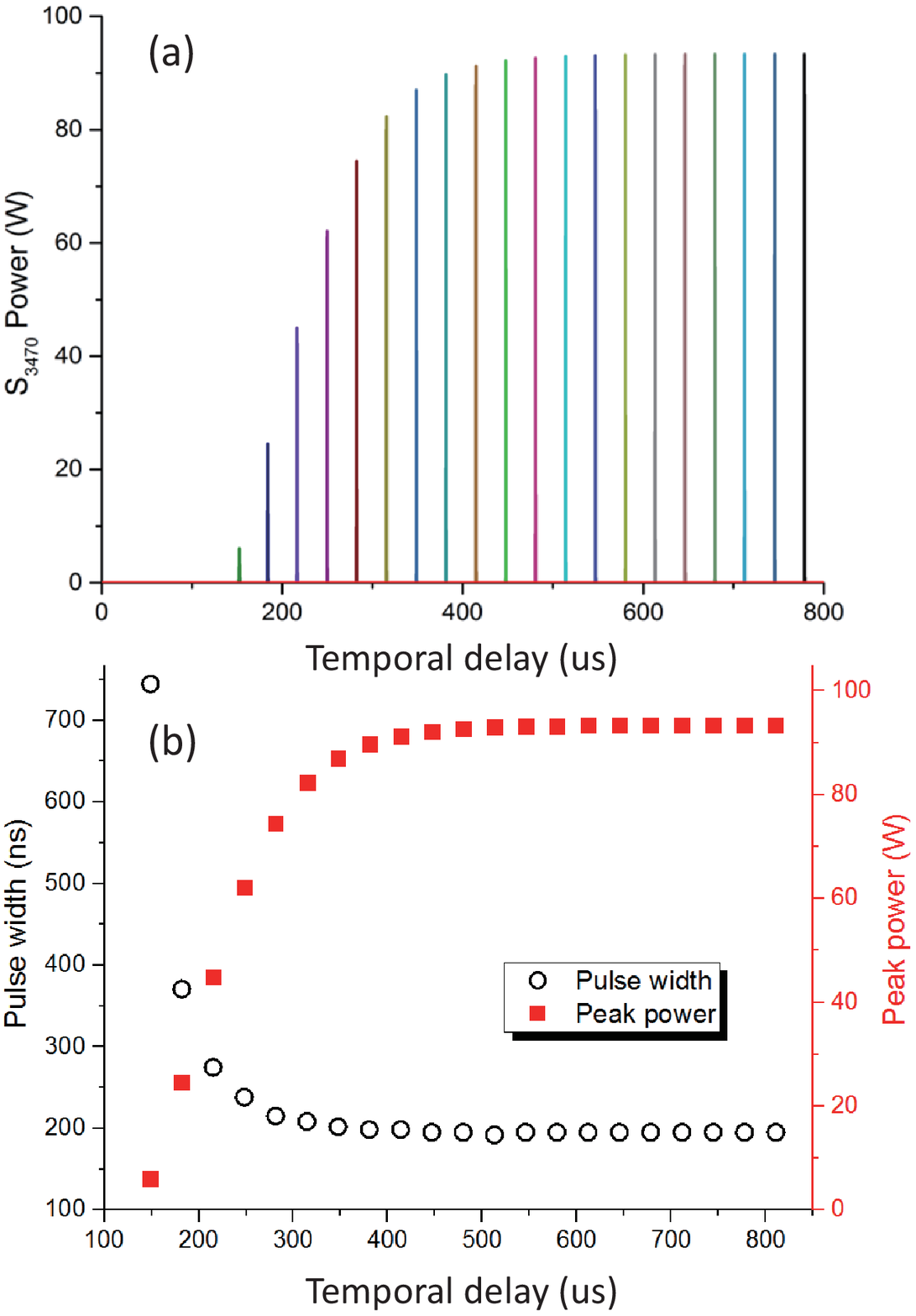}
\caption{Relationship between the temporal delay and the output characteristics. (a) 3.5-$\mu$m pulses generated at different temporal delays. (b) 3.5-$\mu$m pulse width and peak power as functions of the temporal delay.}
\end{figure}
\indent To investigate the relationship between the temporal delay and the 3.5-$\mu$m output, we included more temporal delays in the simulation. Thus we could acquire a graph of the 3.5-$\mu$m pulses generated at different temporal delays as demonstrated in Fig.~4(a). We can see the laser pulses are formed when the temporal delay is larger than 140 $\mu$s. As the delay increases, the laser pulse grows and becomes stable at a delay of around 400 $\mu$s. Fig. 4(b) shows the generated pulse width and peak power as functions of the temporal delay. As the delay increases, the peak power increases while the pulse widths decrease. The stable-outputted pulse width is around 190 ns and the peak power is around 93 W. \\
\indent The results above show the stable output of the 3.5-$\mu$m gain-switched pulses requires adequate accumulation of the $^4I_{11/2}$ level, which suggests the consumption of the population by the 1970-nm pulses will directly impact the 3.5-$\mu$m output: when the 1970-nm pulse energy or the repetition rate is too high, the stable pulsed lasing at 3.5 $\mu$m cannot be formed. We will investigate this phenomenon in the analysis of the cascade-gain-switching below.
\subsection{Cascade gain-switching: single shot regime}
The gain-switched fiber lasers usually output repetitive pulses with repetition rates of several kilohertz to hundred of kilohertz. This repetitive operation has major impact on the output performance of the laser because of the pulse energy accumulation effect as demonstrated in \cite{yan15}. To isolate the influence of the repetitive pulses with other laser parameters, i.e., CW pump power and peak power of pulsed pump, we divided our analysis of the cascade-gain-switching into single shot regime and repetitive pulse regime. For saving computation time, we employed the population distribution of all energy levels of Er$^{3+}$ ions at the time of the 975-nm CW pump was turned on for 1 ms. \\
\begin{figure*}[!!!htb]
\centering
\includegraphics[width=16cm]{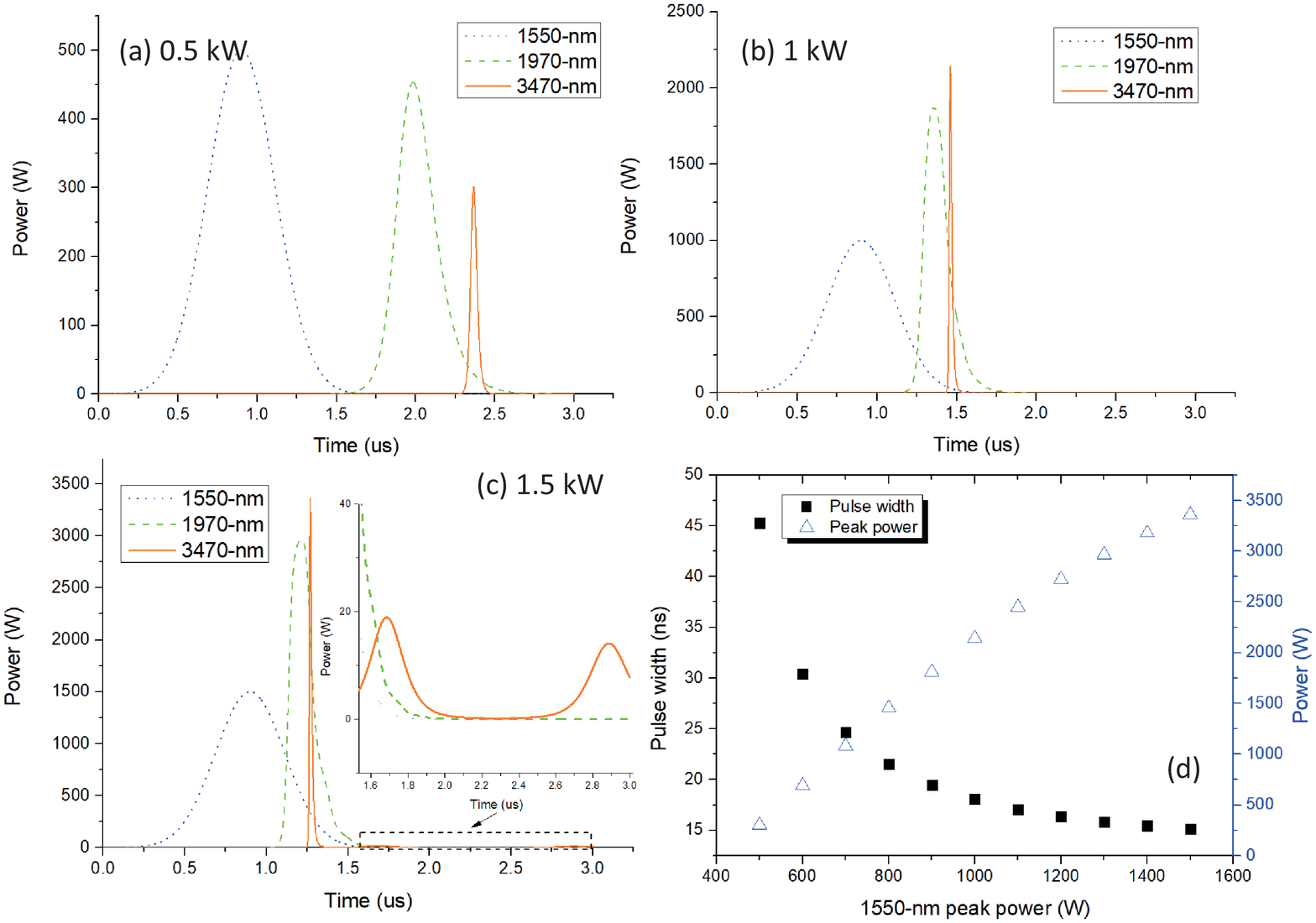}
\caption{Impact of 1550-nm peak power on generated 1970-nm and 3470-nm pulses. 1550-nm peak power: (a) 500 W. (b) 1 kW. (c) 1.5 kW. (d) 3470-nm pulse width and peak power as functions of the 1550-nm peak power.}
\end{figure*}
\indent First, we explored the influence of the 1550-nm peak power on the cascade-gain-switching process. A pulse width of 500 ns was set for the 1550-nm pulse. We employed a CW pump power of 5 W. The results are shown in Fig.~5. In Fig.~5(a), when the pump peak power is 500 W, the generated 1970-nm pulse has a temporal delay of $\sim1.2$ $\mu$s, which can be attributed to the long establishing time of the population on the $^3F_4$ level because of the low pump energy. The 3470-nm pulse, on the other hand, is generated on the trailing edge of the 1970-nm. It can be related to the fact that the life time of the $^4I_{9/2}$ level (8 $\mu$s) is much shorter than that of the $^4F_{9/2}$ level (177 $\mu$s), so the inverted population can be quickly built. The shorter intra-cavity life time of 3470-nm photons also leading to a much shorter pulse width at this wavelength, which is favorable for high-peak power pulse applications.\\
\indent As the 1550-nm pump peak power increases to 1 kW [Fig.~5(b)], the delay between the pump pulse and the generated pulses decrease.  The 1970-nm pulse has a temporal delay of 700 ns and the pulse width is shorten from 650 ns for the 500-W pump to 180 ns for the 1-kW pump. The influence of the increased pump pulse energy is not limited in time domain, the peak powers of the 1970-nm pulse and 3470-nm pulse also increase to around two times higher than that of the pump pulse. The conversion from low peak power to high peak power and from wide pulse to short pulse demonstrates the effectiveness of this cascade-gain-switching approach.\\
\indent When the pump peak power further increases to 1.5 kW [Fig.~5(c)], the changes on the pulse widths of the generated 1970-nm and 3470-nm pulses are neglectable while the peak powers keep increasing, although the peak power of the generated pulses are still around two times higher than that of the pump pulse. So increasing the pump pulse energy to this level can contribute to linearly scale up the peak power but cannot shorten the pulse width any more. It also should be noted that some trailing spikes appear on the generated 3470-nm pulse but the 1970-nm pulse is still free from trailing spikes. It is because the lifetime of the $^3F_4$ level is around two times longer than that of the $^4F_{9/2}$ level, which means more energy can be stored and emitted once other than emitted several times.\\
\indent To further reveal the relationship between the pump peak power and the generated 3470-nm pulse width and peak power, we plot the corresponding functions in Fig.~5(d). As the pump peak power increase from 500 W to 700 W, the 3470-nm pulses are linearly narrowed from 46 ns to 25 ns. When the pump peak power further increases, the rate of pulse shorten decreases. As a pump peak power of 1.3 kW, the 3470-nm pulse width is $\sim17$ ns. The further narrowing of the pulse is neglectable as the pump peak power further increases. The excess pump energy will be released in the form of the trailing spikes. The 3470-nm pulse peak power, on the other hand, shows a better linear trend as the pump peak power increases, even though the increase rate becomes smaller when the pump peak power is larger than 900 W. Increasing the pump peak power from 500 W to 1.5 kW can efficiently scales up the 3470-nm laser peak power from $\sim$ 400 W to 3.4 kW. The effective scaling of the peak power is because a high pump peak power can lead to a high instantaneous population inversion thus generate high laser peak power. The pulse width, on the other hand, is related to many other factors such as the pump pulse energy and the upper-level lifetime, so its relationship with the pump peak power is quite nonlinear. When the pump peak power increases from 500 W to 1 kW, the generated pulses are rapidly narrowed from $\sim45$ ns to $\sim17$ ns. When the pump peak power is larger than 1 kW, increasing the pump peak power can not shortens the 3470-nm pulse any more, the pulse width keeps nearly constant at $\sim16$ ns. \\
\begin{figure}[!!!htb]
\centering
\includegraphics[width=8cm]{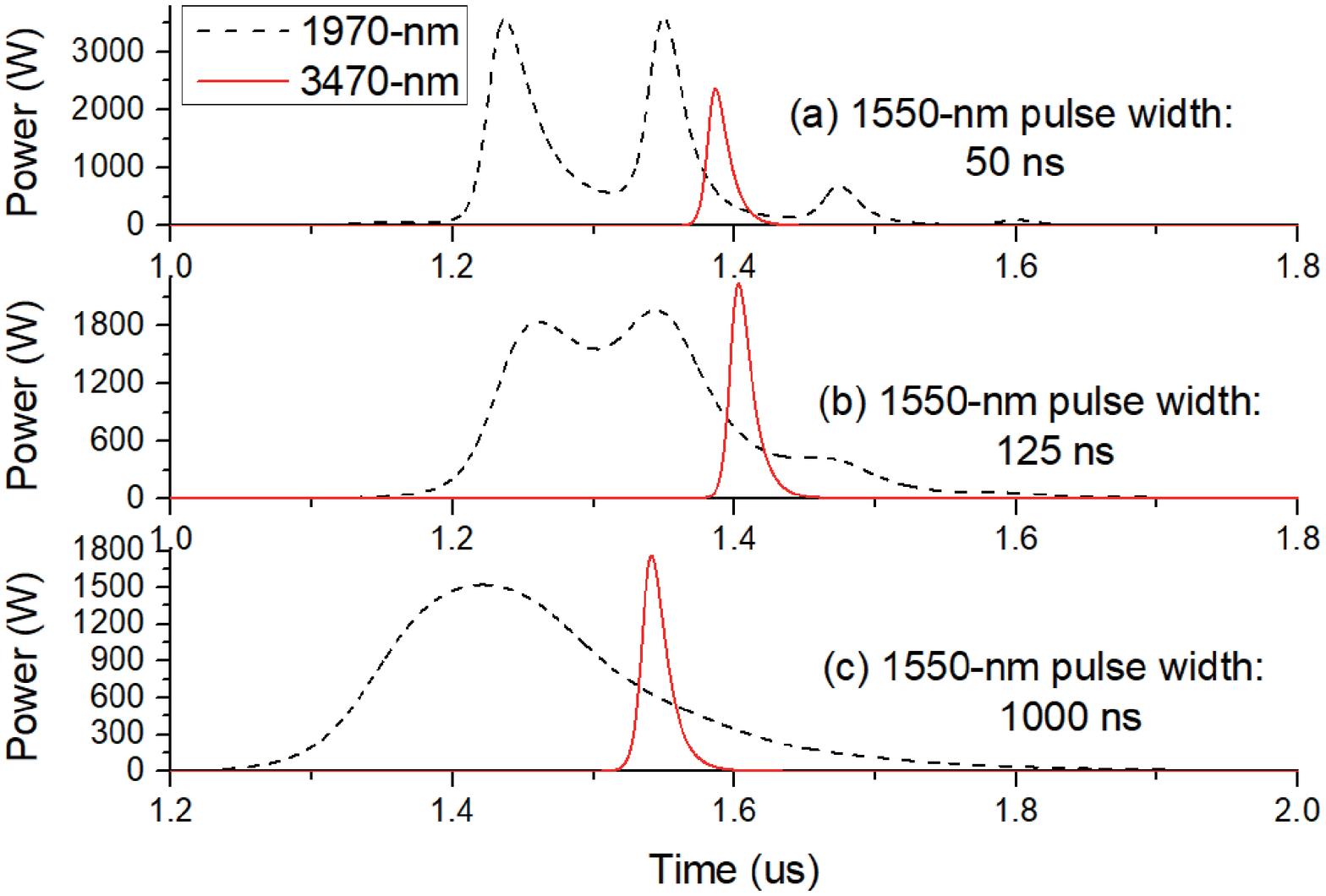}
\caption{Impact of the 1550-nm pump pulse width on the generated pulses. Pump pulse width: (a) 50 ns, (b) 125 ns, (c) 1000 ns.}
\end{figure}
\indent It is also interesting to see how the pulse width of the 1550-nm pump pulse influences the output performance. Because the lasing threshold of gain-switched fiber lasers is directly related to the pump pulse energy \cite{yang13}, for uniformity, we employed a 1550-nm pump pulse energy of 500 $\mu$J for the pump pulses with different pulse widths. In Fig. 6. we demonstrate the intermediate 1970-nm pulses and the generated 3470-nm pulses at the pump pulse widths of 50 ns (a), 125 ns (b), and 1000 ns (c). We can first notice that the generated 1970-nm pulses have significant variation on pulse shape: when the pump pulse has a pulse width of 1000, the 1970-nm has a Gaussian-like pulse shape. When the pump pulse width is 125 ns, some distortions appear on the envelope of the generated 1970-nm pulse. When the pump pulse is further narrowed to 50 ns, the 1970-nm pulse is split into three sub pulses. The transformation of the 1970-nm pulses is related the lasing dynamics of the in-band-pumped Tm-doped fiber lasers. Because no relaxation exists between the pump absorption level and lasing level, the high peak power at short pulse width will lead to rapid lasing. But the pump energy can not be fully consumed by the first lasing pulse, so subsequent pulses will be generated. On the other hand, the pulse shape and width of the 3470-nm pulses are quite similar. It can be attributed to the characteristics of the intermediate 1970-nm pulses, which have similar pulse energy and peak power even though their pulse shape are quite different. So the generation process of the 3470-nm pulses should be similar. Since the 1970-nm pulses are only intermediate product, we can conclude the impact of the pulse width of the pump 1550-nm pulse on the generated 3470-nm pulse can be neglected.\\
\begin{figure}[!!!htb]
\centering
\includegraphics[width=8cm]{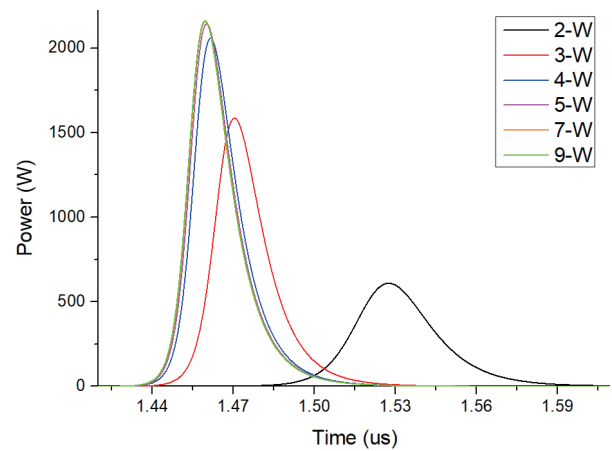}
\caption{Influence of the 975-nm CW pump power on the generated 3470-nm pulses.}
\end{figure}
\indent In the hybrid-pumping scheme, the 975-nm CW pump is employed to stimulate the population from the ground level $^4$I$_{15/2}$ to the $^4$I$_{11/2}$ level, which can be seen as the new ground level for the gain-switching process. It's population distribution will have influences on the lasing dynamics of the 3470-nm pulse, so we also investigated this effect by changing the 975-nm CW pump power to different levels. The 1550-nm pump pulse used here has a pulse width of 500 ns and a peak power of 1 kW. The results are shown in Fig.~7. The pulses generated by different CW pump power are marked by different colors. As the power increases from 2 W to 4 W, the peak power of the generated 3470-nm pulses increases while the pulse width decrease. When the CW pump power is larger than 4 W, the changes on both the peak power and the pulse width can be neglected.  
\subsection{Cascade gain-switching: repetitive pulse regime}
\indent The complex transition dynamics of the cascade gain-switching fiber lasers have major impact on the output performance of the 3.5-$\mu$m pulsed lasing, which has been indicated by the single-pulse analysis above. In the repetitive pulse regime, successive 1550-nm pump pulses will change the temporal features of building and consuming the inversion population. It is interesting to see how these changes influence the 3.5-$\mu$m laser output. Besides, the repetitive operation can effectively demonstrate how this kind of lasers works in the real-world applications. We employed a CW pump power of 5 W, a 1550-nm pulse width of 500 ns and a peak power of 1 kW.\\
\begin{figure}[!!!htb]
\centering
\includegraphics[width=9cm]{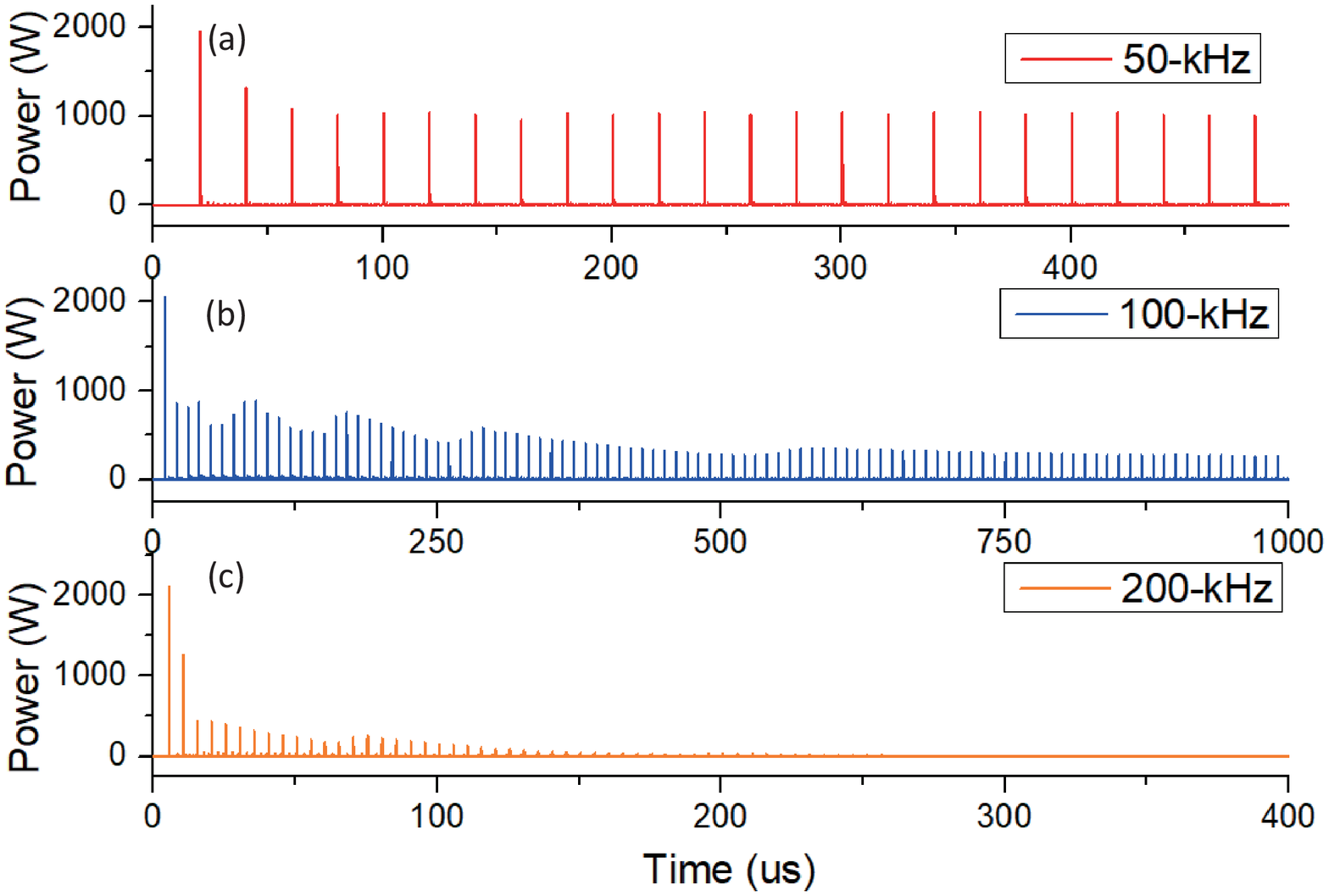}
\caption{Output of the 3.5-$\mu$m laser at the pump pulse repetition rates of (a) 50 kHz, (b) 100 kHz, and (c) 200 kHz.}
\end{figure}
\indent We simulated the 3.5-$\mu$m laser output at different pump pulse repetition rates. The representative results are shown in Fig.~8. At a repetition rate of 50 kHz, stable pulse train output can be achieved less than 100 $\mu$s after few pulses depletes the initial accumulation of the pump energy. At a repetition rate of 100 kHz, strong oscillation of the pulse-to pulse intensity appears. The 3470-nm pulse train turns to be stable when the operation duration is larger than 700 $\mu$s. At a repetition rate of 200 kHz, the stable output pulse trains can no longer be acquired. The intensities of the generated 3470-nm pulses gradually decrease. After $\sim$250 $\mu$s, the cascade gain-switching system stops lasing. We found the stable repetitive output of the 3.5-$\mu$m laser can only be acquired when the repetition rate is $<=$ 100 kHz. This phenomenon should be related to the fact that the depletion rate of the inversion population is fast than the building rate when the repetition rate is too high.  To validate this speculation, we further checked the temporal evolution of the $^4$I$_{11/2}$ level (N$_3$) population at the above repetition rates as demonstrated in Fig.~9.\\
\begin{figure}[!!!htb]
\centering
\includegraphics[width=9cm]{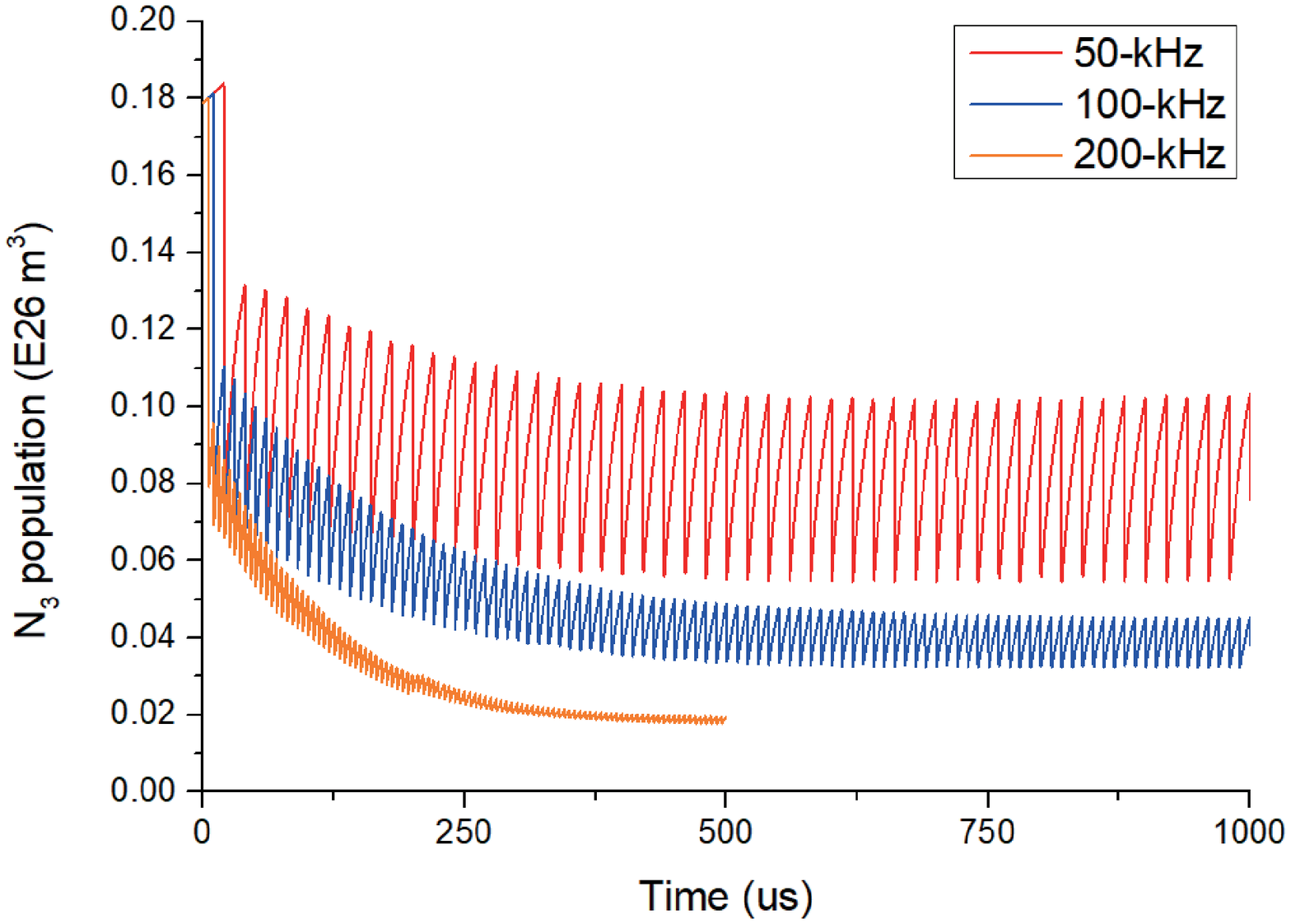}
\caption{Temporal evolution of the $^4$I$_{11/2}$ level (N$_3$) population at the pump pulse repetition rates of (a) 50 kHz, (b) 100 kHz, and (c) 200 kHz.}
\end{figure}
\indent As shown in this figure, at the beginning of the temporal axis, where the 975-nm CW power has already been turned on for 1 ms, so the initial population is $\sim 0.18\times 10^{26}$ m$^3$. For all three kinds of repetition rates, the first few pulses depletes the population of this level to around $\sim 0.08\times 10^{26}$ m$^3$. After that, because the pulse interval of the 50-kHz pump train is larger than others, the population can be recovered to relatively high values by the CW pump and the 1970-nm gain-switched pulses. Then its population stably oscillates in consistent with the pump repetition rate. On the other hand, the 100-kHz pump train prevents the population going back to high levels because of the fast depletion. The envelope of the population gradually becomes smaller. Fortunately, the balance between the consumption and establishment of the population still can be achieved, which leads to the stable oscillation of the population at the same repetition rate after 700 $\mu$s. For the 200-kHz case, the envelope of the population oscillation and the absolute population value keep decreasing until the laser stops working. The population evolution show excellent consistency with the characteristics of the 3.5-$\mu$m laser outputs. It confirms the cascade-gain-switching fiber lasers proposed here have an upper limit for the operation repetition rate.   
\section{Conclusion}
In conclusion, we have proposed a cascade-gain-switching technique for generating 3.5 $\mu$m nanosecond lasers. It has two major aspects of advantages. (1) Compared with the state-of-the-art mid-infrared laser techniques introduced in Section 1, this kind of lasers does not require expensive special-wavelength pump sources and Q-switching/mode-locking components, components used in this configuration are from standard telecommunication industry except the ZBLAN fibers. They have been developed and massively used for more than 30 years thus are low cost and durable. (2) Gain-switched fiber lasers genuinely have a monolithic structure, which is advantageous over the Q-switching and mode-locking components only having a fiber-pigtail packaging. The benefits of the monolithic structure are obvious. It enables the lasers can be stably operated at versatile environments and assembled into compact industrial products. Proof-of-concept numerical simulations have been conducted in this paper. The results shows nanosecond 3.5-$\mu$m pulses with kilowatt peak power can be efficiently generated from this novel configuration. We have also investigated the influence of different pump conditions on the output performance, which can provide guidance for the experimental realization in the future. We also found this kind of lasers had an upper limit for operation repetition rate at around 100 kHz. We think this is a promising approach for generating mid-infrared nanosecond pulses having a wavelength longer than 3 $\mu$m and may be employed in various applications in the future.


\begin{thebibliography}{99}
\bibitem{rudy14} C. W. Rudy, ``Mid-IR Lasers: Power and pulse capability ramp up for mid-IR lasers,'' Laser Focus World 50, 63 (2014). 
\bibitem{np12} ``Mid-infrared photonics,'' Nature Photonics 6, 407-498 (2012).
\bibitem{zhu17}Xiushan Zhu, Gongwen Zhu, Chen Wei, Leonid Vasilyevich Kotov, Junfeng Wang, Minghong Tong, Robert A. Norwood, and N. Peyghambarian, ``Pulsed fluoride fiber lasers at 3  μm [Invited],'' J. Opt. Soc. Am. B 34, A15-A28 (2017).
\bibitem{tokita11} S. Tokita, M. Murakami, S. Shimizu, M. Hashida, and S. Sakabe, ``12 W Q-switched Er:ZBLAN fiber laser at 2.8 μm,'' Opt. Lett. 36, 2812–2814 (2011).
\bibitem{gorjan11} Martin Gorjan, Rok Petkovšek, Marko Marinček, and Martin Čopič, ``High-power pulsed diode-pumped Er:ZBLAN fiber laser," Opt. Lett. 36, 1923-1925 (2011).
\bibitem{li12} Jianfeng Li, Tomonori Hu, and Stuart D. Jackson, "Dual wavelength Q-switched cascade laser," Opt. Lett. 37, 2208-2210 (2012).
\bibitem{hu12} Tomonori Hu, Darren D. Hudson, and Stuart D. Jackson, ``Actively Q-switched 2.9 μm Ho3+Pr3+-doped fluoride fiber laser,'' Opt. Lett. 37, 2145-2147 (2012).
\bibitem{hs14} Ori Henderson-Sapir, Jesper Munch, and David J. Ottaway, "Mid-infrared fiber lasers at and beyond 3.5 μm using dual-wavelength pumping," Opt. Lett. 39, 493-496 (2014).
\bibitem{maes17} Frédéric Maes, Vincent Fortin, Martin Bernier, and Réal Vallée, ``5.6  W monolithic fiber laser at 3.55  μm,'' Opt. Lett. 42, 2054-2057 (2017).
\bibitem{malouf16} Andrew Malouf, Ori Henderson-Sapir, Martin Gorjan, David J. Ottaway, ``Numerical Modeling of 3.5 μm Dual-Wavelength Pumped Erbium-Doped Mid-Infrared Fiber Lasers,'' IEEE Journal of Quantum Electronics 52, 11 (2016).
\bibitem{yang13} Jianlong Yang, Yulong Tang, and Jianqiu Xu, ``Development and applications of gain-switched fiber lasers [Invited],'' Photon. Res. 1, 52-57 (2013).
\bibitem{jiang07} Min Jiang and Parviz Tayebati, "Stable 10 ns, kilowatt peak-power pulse generation from a gain-switched Tm-doped fiber laser," Opt. Lett. 32, 1797-1799 (2007).
\bibitem{creeden08} Daniel Creeden, Peter A. Ketteridge, Peter A. Budni, Scott D. Setzler, York E. Young, John C. McCarthy, Kevin Zawilski, Peter G. Schunemann, Thomas M. Pollak, Evan P. Chicklis, and Min Jiang, "Mid-infrared ZnGeP2 parametric oscillator directly pumped by a pulsed 2 μm Tm-doped fiber laser," Opt. Lett. 33, 315-317 (2008).
\bibitem{swiderski13} Jacek Swiderski, Maria Michalska, and Gwenael Maze, "Mid-IR supercontinuum generation in a ZBLAN fiber pumped by a gain-switched mode-locked Tm-doped fiber laser and amplifier system," Opt. Express 21, 7851-7857 (2013).
\bibitem{yang15} Jianlong Yang, Yao Wang, Geng Zhang, Yulong Tang, and Jianqiu Xu, “High-power highly-linear-polarized nanosecond all-fibre MOPA at 2040 nm,” IEEE Photonics Technology Letters 27(9) 986-989 (2015).
\bibitem{yang132} Jianlong Yang, Yulong Tang, Jianqiu Xu, “Hybrid-pumped, linear-polarized, gain-switching operation of a Tm-doped fibre laser,” Laser Physics Letters 10(5) 055104 (2013).
\bibitem{yang14} Jianlong Yang, Hongqiang Li, Yulong Tang, Jianqiu Xu, “Temporal characteristics of in-band-pumped gain-switched thulium-doped fibre lasers” Journal of the Optical Society of America B 31(1) 80-86 (2014).\bibitem{maes172} Frédéric Maes, Vincent Fortin, Martin Bernier, Réal Vallée, "Quenching of 3.4 $\mu$m Dual-Wavelength Pumped Erbium Doped Fiber Lasers", IEEE Journal of Quantum Electronics 53,1-8 (2017).
\bibitem{hs16} Ori Henderson-Sapir, Jesper Munch, and David J. Ottaway, "New energy-transfer upconversion process in Er3+:ZBLAN mid-infrared fiber lasers," Opt. Express 24, 6869-6883 (2016).
\bibitem{yang12} J. Yang, Y. Tang, R. Zhang, and J. Xu, ``Modelling and characteristics of gain-switched diode-pumped Er-Yb codoped fibre lasers,'' IEEE J. Quantum Electron. {\bf 48}(12), 1560--1567 (2012).
\bibitem{yang152} J. Yang, Y. Wang, G. Zhang, Y. Tang, and J. Xu, ``Influences of pump transitions on thermal effects of multi-kilowatt thulium-doped fibre lasers,'' arXiv. 1503.07256, (2015).
\bibitem{yang16} Jianlong Yang, Haizhe Zhong, Shuaiyi Zhang, Dianyuan Fan, ``Theoretical Characterization of the Ultra-Broadband Gain Spectra at ∼1600–2100 nm From Thulium-Doped Fiber Amplifiers,'' IEEE Photonics Journal 8, 1400310 (2016).
\bibitem{yan15} Shuo Yan, Yao Wang, Yan Zhou, Nan Yang, Yue Li, Yulong Tang, and Jianqiu Xu, "Developing high-power hybrid resonant gain-switched thulium fiber lasers," Opt. Express 23, 25675-25687 (2015).
\end{thebibliography}
\end{document}